# Characterization of Micro-Roughness Parameters and Optical Properties of Obliquely Deposited $HfO_2$ Thin Films

**R B Tokas*, S Jena, P Sarkar, S Thakur, and N K Sahoo**

Atomic & Molecular Physics Division, Bhabha Atomic Research Centre,

Trombay, Mumbai-400085, India

*Email: tokasstar@gmail.com

**Abstract:** Oblique angle deposited oxide thin films have opened up new dimensions in fabricating optical interference devices with tailored refractive index profile along thickness by tuning its microstructure by varying angle of deposition. Microstructure of thin films strongly affects surface morphology as well as optical properties. Since surface morphology plays an important role for the qualification of thin film devices for optical or other applications, it is important to investigate morphological properties. In present work, $HfO_2$ thin films have been deposited at several oblique angles. Morphological statistical parameters of such thin films viz., correlation length, intrinsic roughness, fractal spectral strength, etc., have been determined through suitable modelling of extended power spectral density function. Intrinsic roughness and fractal spectral strength show an interesting behaviour with deposition angle and the same has been discussed in the light of atomic shadowing, re-emission and diffusion of ad-atoms. Further refractive index and thickness of such thin films have been estimated from transmission spectra. Refractive index and grain size depict an opposite trend with deposition angle and their variation has been explained by varying column slanting angle and film porosity with deposition angle.

## 1. Introduction

Oblique angle deposition has been attracting researchers due to its applications in interference devices, micro sensors, microelectronics, photonic crystals, and rugate structures based devices. Now-a-days, it is being used for fabricating precision interference filters [1, 2] in which refractive index is varied by varying the angle of deposition resulting in varying porosity due to atomic shadowing and limited ad-atom diffusion [3, 4] during growth. It generally works at angles greater than 60° with normal to the substrate. When the angle reaches around 80°, it is termed as glancing angle deposition (GLAD). Oblique angle deposition results in special nano and microstructure of thin films. By employing substrate rotation and varying deposition angle, different geometries like pillar, helix, zigzag, erect columns etc. have been achieved successfully [5-8]. Zhu et al. have fabricated multi-stop band interference rugate filter exploiting GLAD technique [9]. Park et al. have fabricated bilayer circular filter by GLAD deposition of $TiO_2$ thin films [10]. Fahr et al. have developed optical rugate filters for light trapping in solar cells[11]. Gasda et al. have fabricated nono-rod proton exchange membrane fuel cell cathodes by glancing angle deposition of Carbon[12]. Researchers have also developed GLAD magnetic data storage device[13], antireflection coating[14], selective polarization transmission filter[15], narrow band pass rugate filter [16], and relative humidity sensors[17] by employing oblique angle deposition.

It is well understood that surface morphology affects the functionality of thin film and multilayer devices for optical and other applications [18]. Surface morphology strongly perturbs the amount and distribution of scattered light from optical components and such scattering is a performance limiting factor for optical devices. Hence, it is of high importance to characterize micro-roughness parameters of such obliquely deposited $HfO_2$ thin films to assess their surface morphological properties. Generally RMS roughness of surface is taken as the parameter to characterize surface morphology. However, surface roughness parameter computed from RMS distribution of heights only does not take in to account the lateral distribution of surface features. Power spectral density function (PSDF) provides more complete description of surface topography. PSDF describes two aspects of surface roughness viz., the distribution of heights from a mean plane, and the lateral distances over which height variations occur [19-21]. Moreover, PSDF also provides useful information on superstructures and fractals of surfaces. Fractal geometry and scaling concept can concisely describe the rough surface morphology [22, 23]. Surface morphology at different scales is believed to be self-similar and related to the fractal geometry. Fractal analysis can extract many different kind of information from measured surface morphology and that makes fractal approach a very attractive and useful in describing surface statistics of thin films.

Hafnium oxide exhibits high refractive index, high band gap [20, 24], high laser induced damage threshold [25-27] and transparency from ultraviolet to mid-infrared (0.20-10μm) [28]. It is widely used high index coating material for fabrication of multilayer interference devices. In present work, 2-D extended PSDF has been computed from measured AFM data for entire set of obliquely deposited $HfO_2$ thin films by combining PSDF of three different scan sizes. Different PSD models in combination have been fitted with the computed PSDF to extract fractal parameters, correlation length, intrinsic RMS roughness and contribution of aggregates or superstructure to surface roughness. Further, the refractive index and film thicknesses have been computed from transmission measurements. We have found very interesting correlations among micro-roughness parameters, refractive index and angle of deposition.

**2. Experimental detail**

In present work, $HfO_2$ thin films were deposited at angles (α) = 0°, 40°, 50°, 57°, 62° 68° and 80°. The samples have been designated as S-7, S-6, S-5, S-4, S-3, S-2 and S-1 respectively. Thin films were deposited on fused silica substrate at 200°C temperature by reactive electron beam (EB) evaporation. Schematic of oblique angle deposition is shown in Fig. 1. Before deposition, entire batch of substrates were cleaned in ultrasonic cleaner and vapour degreaser to achieve good quality films. α was defined as the angle between normal to the substrate plane and incident vapour flux as shown in Fig. 1. Different values of α were set by tilting the substrate whereas direction of the incoming vapour flux was held fix. Distance between substrate and vapour source was kept ~ 45 cm. The base pressure prior to deposition was kept ~$1x10^{-5}$ mbar. During deposition, high purity (99.9%) oxygen was supplied to the deposition chamber through mass flow controller to maintain stoichiometry of $HfO_2$ thin films. Rate of deposition and film thicknesses were monitored and controlled by Inficon make 'XTC2' quartz crystal monitor. An optimized oxygen partial pressure of $1x10^{-4}$ mbar and deposition rate of 5Å/s were maintained during deposition. NT-MDT make P47H AFM system was used for morphological measurements of obliquely deposited $HfO_2$ thin films. A super sharp diamond like carbon (DLC) coated Si probe having tip curvature 1-3 nm, resonance frequency 198 kHz and force constant 8.8 N/m has been used. Length and width of DLC cantilever probe were 125 and 35 μm respectively. DLC coated AFM probe being very hard and anti-abrasive, was chosen to get the consistency in the measurements [29]. Three different measurements having scan sizes, 2.5x2.5, 5x5 and 10x10 $\mu m^2$ with spatial resolution of 512x512 points, have been taken for all the films. For optical characterization, transmission measurements were performed from 300-1200 nm with a wavelength resolution of 1 nm on Shimadzu make UV-VIS-NIR 3101PC spectrophotometer.

## 3. Computation and analysis of power spectral density function

PSD function can be derived from many different measurements such as, morphological measurement by surface profilometer, bi-directional reflectance distribution function and AFM measurement of surface profile [19, 21]. Among all, AFM is widely used and an excellent tool to characterize rough surfaces having height irregularities not more than few microns. There are large numbers of publications which describe surface statistics thoroughly [30-32]. In present paper, we have adopted the formulation described in refs. [33, 34] for the computation of PSDF as following

$$S_2(f_x f_y) = \frac{1}{L^2}\left[\sum_{m=1}^{N}\sum_{n=1}^{N} Z_{mn} e^{-2\pi i \Delta L(f_x m + f_y n)} (\Delta L)^2\right]^2 \tag{1}$$

Here $S_2$ is the 2-dimensional PSDF, $L^2$ is the scanned surface area, N is the number of data points in both X and Y direction of scanned area, $Z_{mn}$ is the surface profile height at position (m,n), $f_x$ & $f_y$ are the spatial frequencies in X and Y directions respectively. ΔL (L/N) is the sampling interval.

Computation of PSDF is further followed by transition to polar coordinates in frequency space and angular averaging (φ)

$$S_2(f) = \frac{1}{\pi}\int_0^{2\pi} S_2(f,\varphi) d\varphi \tag{2}$$

As the PSDF depends only on one parameter, it will be plotted in all our figures as a 'slice' of the 2-D PSDF with unit '(length)$^4$.

PSDF obtained from single AFM scan has roughness in limited band of spatial frequencies and the band width depends on scan area and sampling interval. Artefacts can also constrain frequency band width of PSDF. Fortunately such band width limitation can be eliminated by combing topographical measurement performed on different scan size provided following conditions are satisfied:

(1) Spatial frequency range on which different scan size measurements are performed should overlap partially.

(2) Different PSDF should be of the same order of magnitude in the overlap region.

With the conditions mentioned above, combined PSDF at a frequency is given by geometrical averaging:

$$PSD_{combined}(f) = \left[ \prod_{i=1}^{M} PSD_i(f) \right]^{1/M} \quad (3)$$

Here M is the PSDF overlapping at concerned frequencies.

In present work, PSD functions have been computed separately for scan area, 2.5x2.5, 5x5 and 10x10 μm$^2$ and combined together in a suitable manner taking care of all the conditions mentioned above. Experimental PSDF computed for morphologies of obliquely deposited $HfO_2$ thin films needs appropriate analysis models so that an extensive interpretation of PSDF can provide deep insight of morphological statistical parameters. Several mathematical models alone or in combinations has been proposed and used by researchers to interpret experimental PSDF. The most used extended model for PSDF of thin films is the sum of Henkel transforms of the Gaussian and exponential autocorrelation functions [35, 36]. But such model fails when wide spatial frequency range is considered. To describe roughness over large spectral frequencies, PSD model should comprises contribution from substrate, pure thin film and aggregates or superstructures. PSDF of substrates generally follows inverse power law with spatial frequency (assuming fractal like surfaces) and is given as following [37]

$$PSD_{fractal}(f; K, \gamma) = \frac{K}{f^{\gamma+1}} \quad (4)$$

Here K is spectral strength of fractal and $\gamma$ is fractal spectral indices. This PSD formulation follows self-affine surfaces only and fractal dimension *Fd* is given as following

$$Fd = \frac{4-\gamma}{2} \quad ; 0 < \gamma < 2 \quad (5)$$

When $\gamma = 0$ i.e. $Fd = 2$, surface is extreme fractal, for $Fd = 1.5$, surface is Brownian fractal and for $Fd=1$, surface is marginal fractal. Apart from substrate fractal contribution towards total roughness, thin films also exhibits strong fractal characters especially at higher spectral frequencies. PSDF of pure thin film is conventionally characterized by ABC or k-correlation model [38, 39]:

$$PSD_{ABC} = \frac{A}{(1+B^2 f^2)^{(C+1)/2}} \tag{6}$$

A, B, C are model parameters. Equivalents RMS roughness (contribution from pure thin film) and correlation length which depicts the grain size, are related to A, B and C parameters as following;

$$\sigma_{ABC}^2 = \frac{2\pi A}{B^2(C-1)},\ \tau_{ABC}^2 = \frac{(C-1)^2 B^2}{2\pi^2 C} \tag{7}$$

Models discussed so far are monotonically decreasing function of spatial frequency and do not accounts for any local maxima in PSDF while experimental PSD functions of our films exhibits one or more local maxima due to contributions from aggregates. Such peaks in experimental PSD can be accounted by using Gaussian function with its peak shifted to a non-zero spatial frequency as described in ref. [40]. For our thin films which exhibit one or more peaks in PSD function, we have used the combination of all the three PSD models and the combined formulation is as follows:

$$PSD_{Total} = \frac{K}{f^{\gamma+1}} + \frac{A}{(1+B^2 f^2)^{(C+1)/2}} + \sum_m \pi\sigma_{sh,m}^2 \tau_{sh,m}^2 \exp\left\{-\pi^2\tau_{sh,m}^2 (f - f_{sh,m})^2\right\} \tag{8}$$

Here, $\sigma_{sh}$ and $\tau_{sh}$ correspond to height and size of superstructure. PSD functions of entire set of our obliquely deposited $HfO_2$ thin films have been fitted using above formalism to obtain useful thin film surface statistical parameters.

**4. Determination of optical constants from transmission spectra**

Prior to the determination of optical constants of thin film, substrate transmission spectra was fitted with its theoretical expression [41] using suitable dispersion relation to estimate substrate refractive index and extinction co-efficient. The procedure for deriving refractive index, and thickness of thin films from the measured transmission spectrum is detailed in

ref.[42].Theoretical transmission of single layer thin film was generated using Sellmeier's dispersion model. $\chi^2$ (chi) square minimization [41] has been carried out to determine the fitting parameters. Refractive index (n) and film thickness were computed from the fitting parameters.

**5. Result and discussion:**

Generally PSDF computed from single place AFM scan is very noisy and hence morphological parameters deduced from it may be erroneous. One way to eliminate noise is to carry out data smoothing, which introduces artefacts in data and may lead to wrong estimation of micro-roughness parameters. Other way which we have chosen in present paper is to perform many scans of the same size at different places and then average them. We have performed scans of same size at 8 different places over thin film surface. In Fig. 2, PSDFs of thin film, S-3 computed from equation (1) for scan size 2.5x2.5 $\mu m^2$ at different places are shown along with their average. It is worth to notice that after averaging, fluctuations in PSDF have reduced to a great extent. Extended PSDF for entire set of obliquely deposited $HfO_2$ thin films which has been computed from AFM measurement using equation (1), equation (2) and equation (3), are plotted in Fig. 3. Extended PSDF of thin film S-1, deposited at $80^o$, depicts the highest roughness for full range of spectral frequencies among all the films. Combined PSD model as described by equation (8), has been used to fit the experimental extended PSDF for entire set of thin films. The fitting parameters, intrinsic film roughness and correlation length are listed in Table-1. Experimental and fitted extended PSDF for thin films S-1 and S-7 are shown in Fig. 4(a) and Fig. 4(b) respectively. Fitting quality justifies the use of combined PSDF model. Contribution of different model components to total extended PSDF or spectral roughness has also been computed and plotted in Fig. 4(a) and Fig. 4(b) for films S-1 and S-7 respectively. It can be noted from table-1 that entire set of films except S-1, has been fitted using two shifted Gaussian peaks.

PSDF of film S-1 fits very well using single shifted Gaussian peak only. Fig.5 shows a very interesting trend of intrinsic RMS roughness $\sigma_{ABC}$ and spectral strength of fractal (K) with respect to α. For lower values of α, $\sigma_{ABC}$ is nearly constant (from 0 to 50$^o$). For intermediate angles (50° to 70°), it depicts decreasing trend and then again increases sharply with increase in deposition angle (α ≥ 70°) exhibiting the highest value for film S-1. Such variation of $\sigma_{ABC}$ with α can be explained in the light of atomic shadowing, re-emission and diffusion of ad-atoms. At glancing angle (α ~ 80°) as shown Fig. 1, the velocity component $V\cos\theta$ (*θ = 90-α) of* incoming ad-atoms along the substrate surface is the highest ($\theta$ ~ 10°) and it makes ad-atoms to diffuse on substrate surface and try to smoothen surface. The velocity component $V\sin\theta$ of ad-atom perpendicular to the plane of substrate is the lowest and hence sticking probability of ad-atom to substrate is the lowest at glancing angle. Low sticking probability leads to high re-emission of ad-atoms which gives smoothing effect to the surface of slanted angle columnar growth film [43]. On the other hand, for glancing angle, surface roughening due to atomic shadowing effect is very high and dominates surface smoothing effects due re-emission and diffusion of ad-atoms [43, 44]. Consequently, GLAD film depicts the highest surface roughness amongst all the films. As angle *α* decreases i.e. *θ* increases, shadowing effect tends to diminish very fast [44]. For angle α below 70°, shadowing offers very small roughening effects. However, re-emission of ad-atoms also decreases with angle *θ* due to increase in sticking probability, smoothing due to re-emission and diffusion of ad-atoms starts dominating roughening due to shadowing effect. Consequently, effective smoothing of surface for intermediate angles (70°-57°) occurs and $\sigma_{ABC}$ depicts lower values for intermediate oblique angles. Now as the angle *θ* increases further, atomic shadowing creates negligible roughening effects. Below 50° angle of deposition, slanted columnar growth tends to disappear and straight & dense columnar growth occurs. Smoothing effect also decreases due to the low of re-emission and diffusion of ad-atom for lower angles of deposition.

Consequently, $\sigma_{ABC}$ increases slightly and then saturates as the angle α tends to zero (normal deposition). Spectral strength of fractal (K), which is the measure of strength of fractal components in surface, follows the similar trend as of $\sigma_{ABC}$ with angle α. Such behaviour of K may be the attribute of combined effect of atomic shadowing and re-emission & diffusion of ad-atoms. Similar trend of $\sigma_{ABC}$ and K sets a proportionality relation between $\sigma_{ABC}$ and K. As listed in table-1, correlation length ($\tau_{ABC}$) increases monotonically with angle α, up to α=65°. As the angle α increases further, $\tau_{ABC}$ depicts an abrupt increase. The lowest and the highest value of $\tau_{ABC}$ are 26.7 and 74.6 nm for thin film S-7 & S-1 respectively. $\tau_{ABC}$ represents the size of geometrical grain on the surface. The variation of $\tau_{ABC}$ with angle α is corroborated by 2-D AFM images of thin films as shown in Fig. 6. The highest grain size of film S-1 is the consequence of dominant atomic shadowing effects at glancing angle. Diameter of slanted columns is very high (~ 50-100 nm) at glancing angles which ultimately leads to bigger grain size on surface. For the reference of column diameter, inter-columnar distance and column slanting angles, we have reported cross-sectional FESEM characterization of such films in our earlier work [29]. As the angle α decreases, atomic shadowing effect decreases. This leads to decrease in surface grain size because of decrease in column slanting angle. In Table-1, the values of fractal indices ($\gamma$) and fractal dimensional ($Fd$) are listed. The value of $\gamma$ increases gradually with α and saturates for α ≥ 62°. It can be noted that $\gamma$ varies from 0.33 to 0.48 and hence $Fd$ varies from 1.84 to 1.76. The value of $Fd$ is close to 2 for all the film samples. It indicates that thin film surfaces are close to extreme fractal or between extreme and Brownian fractal. Relatively lower fractal dimension for near glancing angle deposited films indicates that film surfaces are more close to Brownian fractal. The values of $\sigma_1, \tau_1$ & $\sigma_2, \tau_2$ which are roughness contribution and correlation length of aggregates or superstructures for shifted Gaussian peak-1 and peak-2 respectively, are listed in Table-1. It indicates that contribution of aggregates to total PSDF or spectral roughness dominates for lower spatial frequencies and their contribution for higher frequencies is negligible. Aggregates size and their contribution towards spectral roughness are not definite with α and hence no correlation can be set between shifted Gaussian peak parameters and angle α.

Fig.7 (a) presents the transmission spectra of all the thin films. Transmission spectra depict a decrease in visibility of interference fringes with the increase in angle α. Such outcome may be attributed to the increase in porosity in films with α. Again variation in film porosity is governed by film microstructure which changes due to varying atomic shadowing effect with deposition angle. It may also be noted from Fig. 7(a) that for film S-1 (GLAD), the absorption for wavelengths less than 350 nm increases sharply and the same may be the contribution from multiple reflections of light between the columns inside the film and from high diffuse scattering from rough surface [45, 46]. Such loss of light due to scattering or multiple reflections inside thin film become visible and dominant for wavelengths ≤ 300 nm and reason for the same is high inter-columnar distance in GLAD thin film. As the angle $\alpha$ decreases, inter columnar distance reduces steeply and become very small compared to wavelengths of interest. Hence, the films deposited at lower oblique angle do not show any additional absorption. In Fig. 7 (b), experimental & fitted transmission curves are shown for film S-3 ($\alpha=62°$). Suitability of Sellmeier's dispersion model is justified by the fitting quality depicted in Fig. 7(b). Film thicknesses determined through modelling are 622, 487, 390, 478, 398, 426 and 421 nm for films S-1, S-2, S-3, S-4, S-5, S-6, and S-7 respectively. Fig. 8 presents the dispersive values of refractive index computed from modelling of transmission spectra of thin films. Finally, the variation of correlation length and refractive index with deposition angle are plotted together in Fig. 9. Variation of correlation length depicts an opposite trend to refractive index with deposition angle. Such behaviour depicts a strong correlation between grain size and refractive index. Refractive index at wavelength of 600 nm varies between 1.37 and 1.93 as the angle α varies from 80° to 0°. The lowest refractive index of 1.37 is exhibited by GLAD film and is less than the refractive index of fused silica substrate (n=1.45 at λ=600 nm). Consequently, thin film S-1 renders an antireflection effect to fused silica substrate and the same is shown in Fig 7(a). Therefore, it is concluded that variation in microstructure of obliquely deposited $HfO_2$ thin film has great impact on their optical and morphological properties.

## 5. Conclusion

Several $HfO_2$ thin films have been deposited at different oblique angles varying from 0° to 80° by reactive electron beam evaporation. Such thin films posses special microstructures due to atomic shadowing and limited ad-atom diffusion during growth. Varying microstructure with deposition angle also affects morphological and optical properties of thin film. Effect deposition angle on morphological and optical properties has been studied extensively through extended power spectral density function and transmission spectra. Among all the thin films, GLAD film exhibits the highest grain size and intrinsic RMS surface roughness. Intrinsic roughness and fractal spectral strength obtained from the analysis of extended power spectral density function follow the similar tend with deposition angle. Behaviour of surface morphological statistical parameters and refractive index with deposition angles have been explained by the combined effect of atomic shadowing, re-emission of ad-atoms and diffusion of ad-atoms.

## References:-

**Caption of Figures:**

Fig.1: Depicts the schematic of oblique angle deposition and mechanism for slanted columnar growth due to shadowing effects.

Fig. 2: Shows the reduction of noise in PSDF by averaging of PSDF of scans size 2.5 x 2.5 µm measured at 8 different places for film S-3

Fig. 3: Displays extended PSDF for all the obliquely deposited $HfO_2$ thin films.

Fig. 4(a), Fig 4(b): Depicts experimental and fitted extended PSDF of sample S-1 & S-3 respectively. Fitted PSDF has also been de-convolution in different model components.

Fig. 5: Variation of intrinsic RMS roughness and fractal spectral strength with angle of deposition for all the $HfO_2$ thin films

Fig. 6: 2-D AFM images depicting the grain size of obliquely deposited thin films.

Fig. 7(a): Shows the measured transmission spectra for entire set of obliquely deposited $HfO_2$ thin films.

Fig. 7(b): Measured and fitted transmission curves for thin film S-3. Fitting was carried out using Sellmeier's dispersion model.

Fig. 8: Plots of dispersive values of refractive index for all the thin films.

Fig. 9: Variation of refractive index and correlation length with angle of deposition.

**Fig. 1**

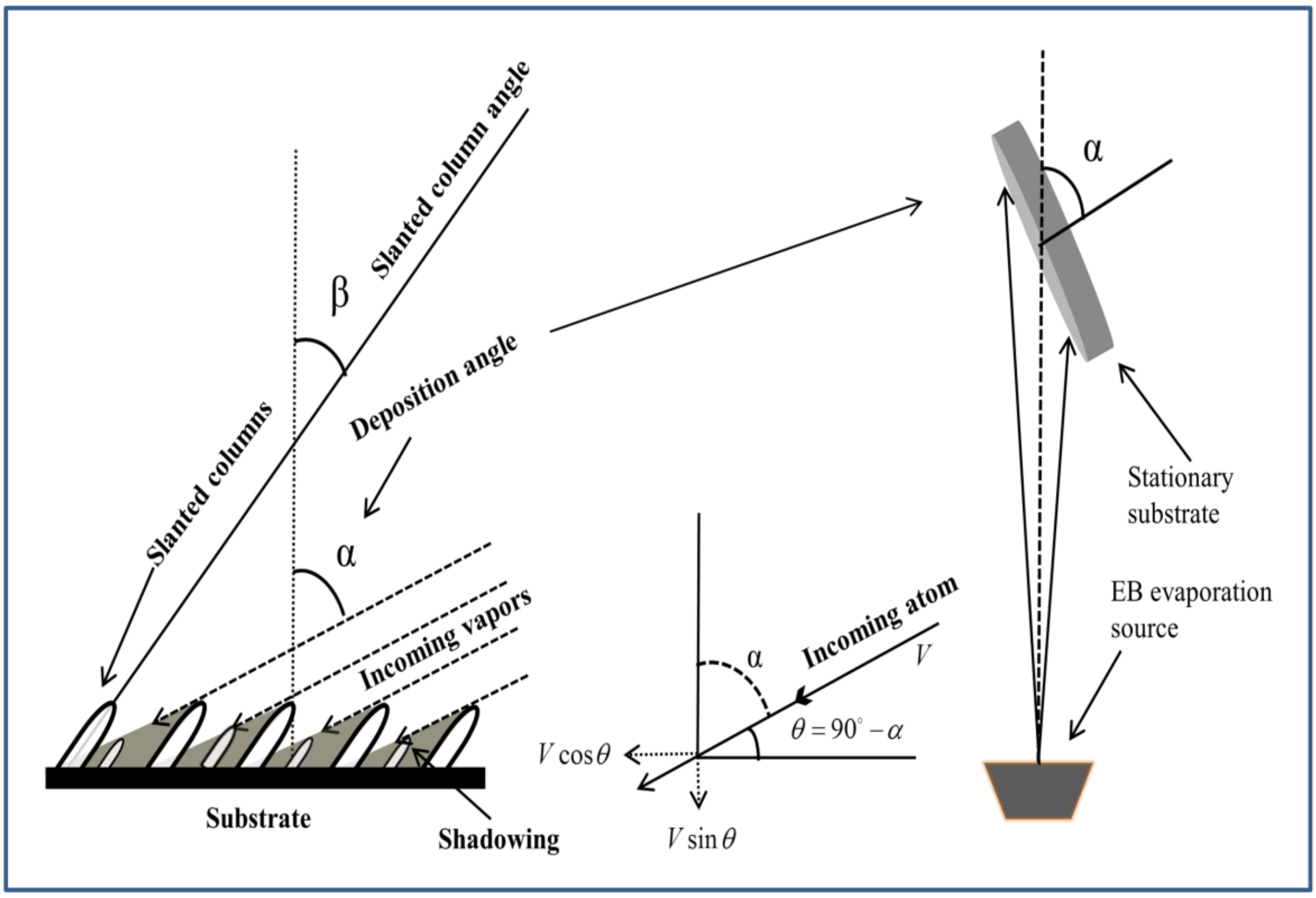

**Fig. 2:**

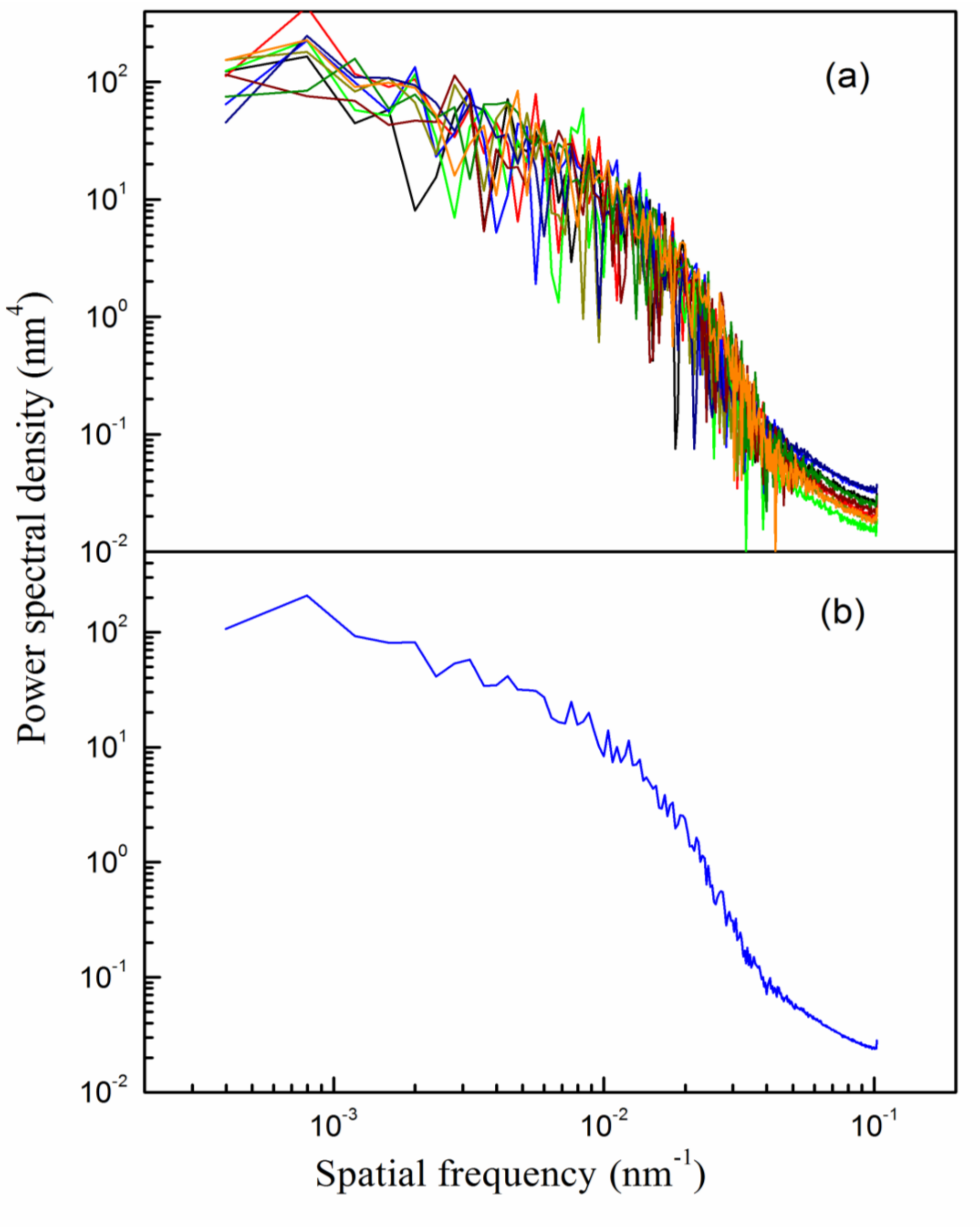

**Fig. 3**

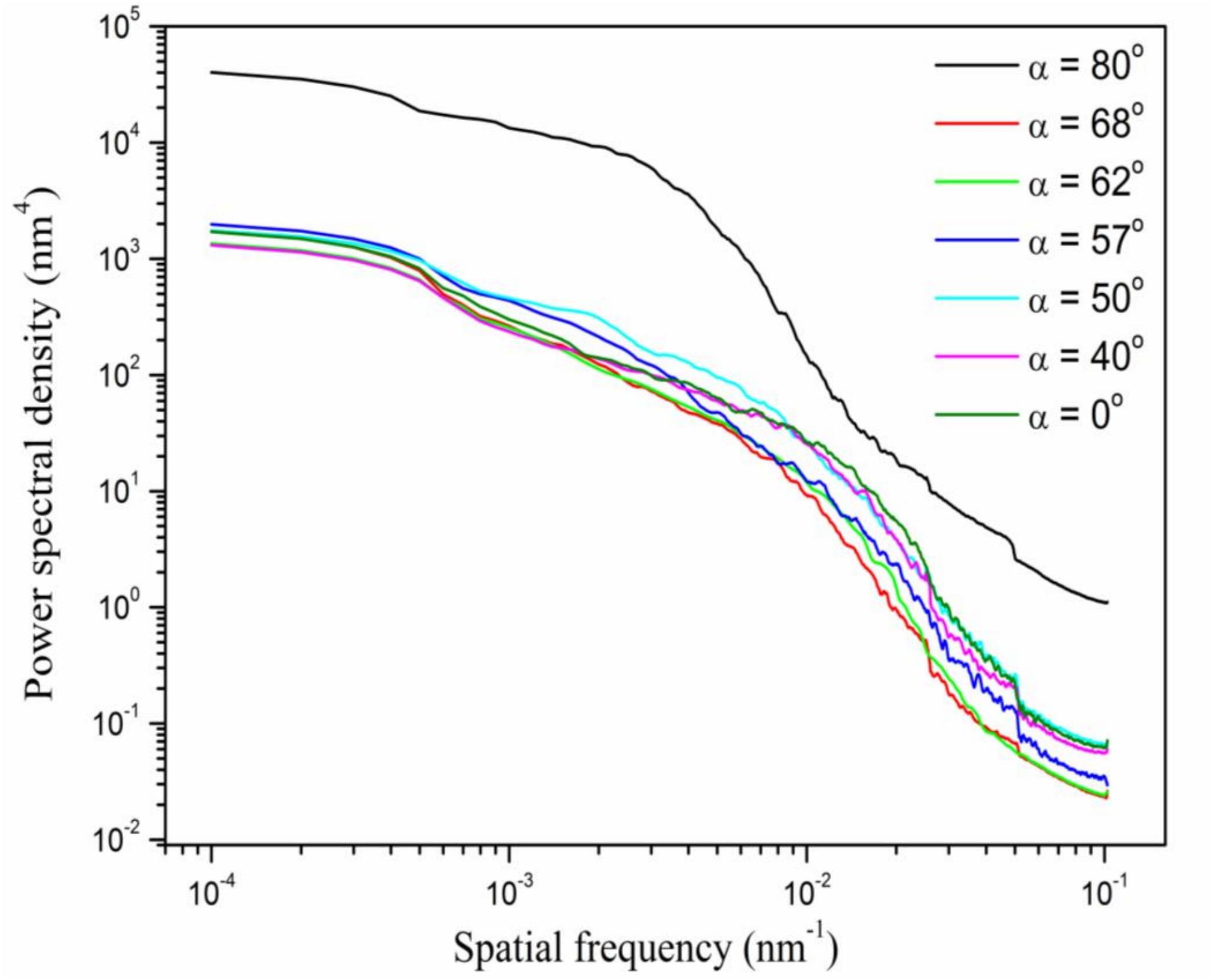

**Fig. 4(a):**

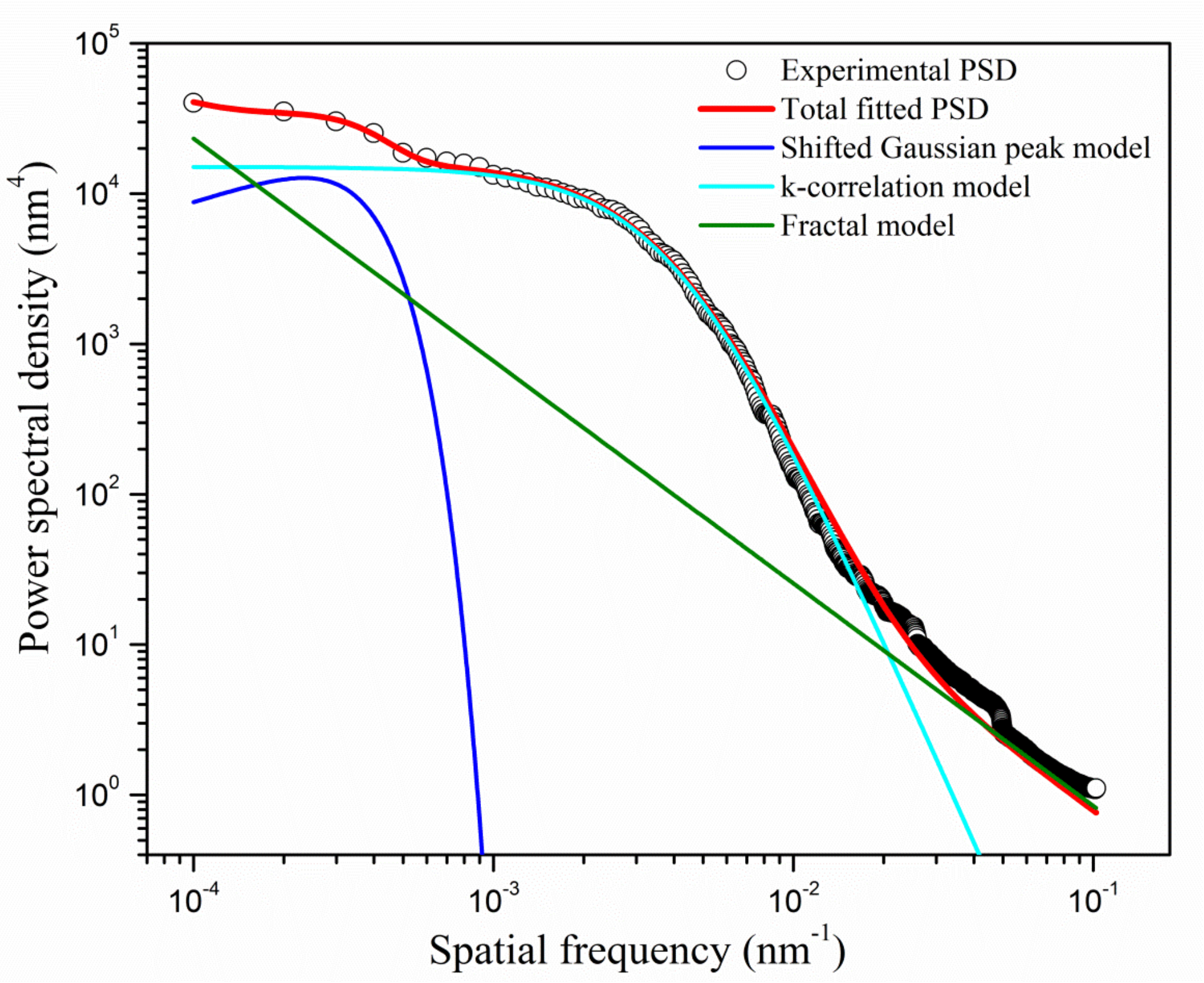

**Fig. 4(b)**

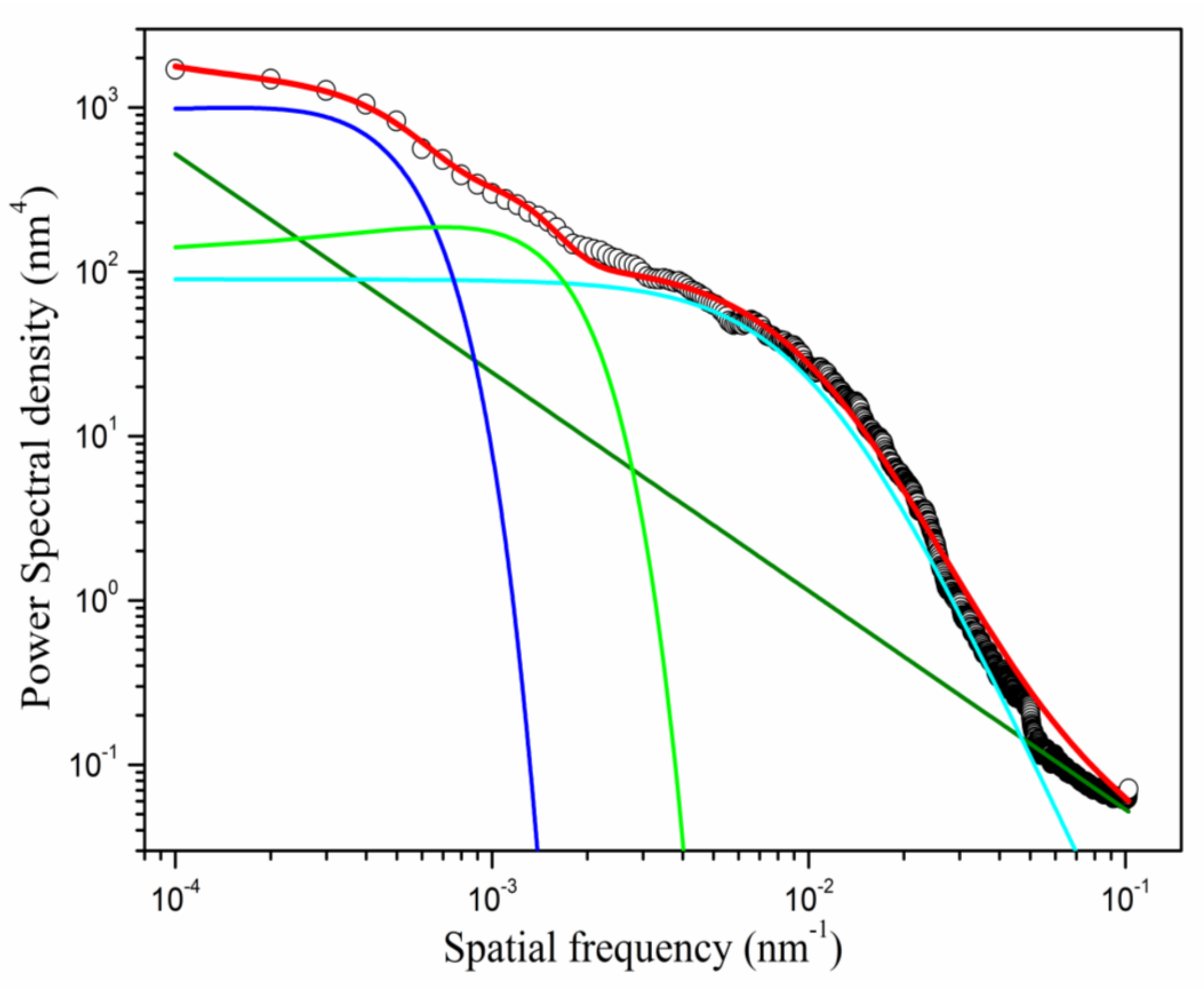

**Fig. 5**

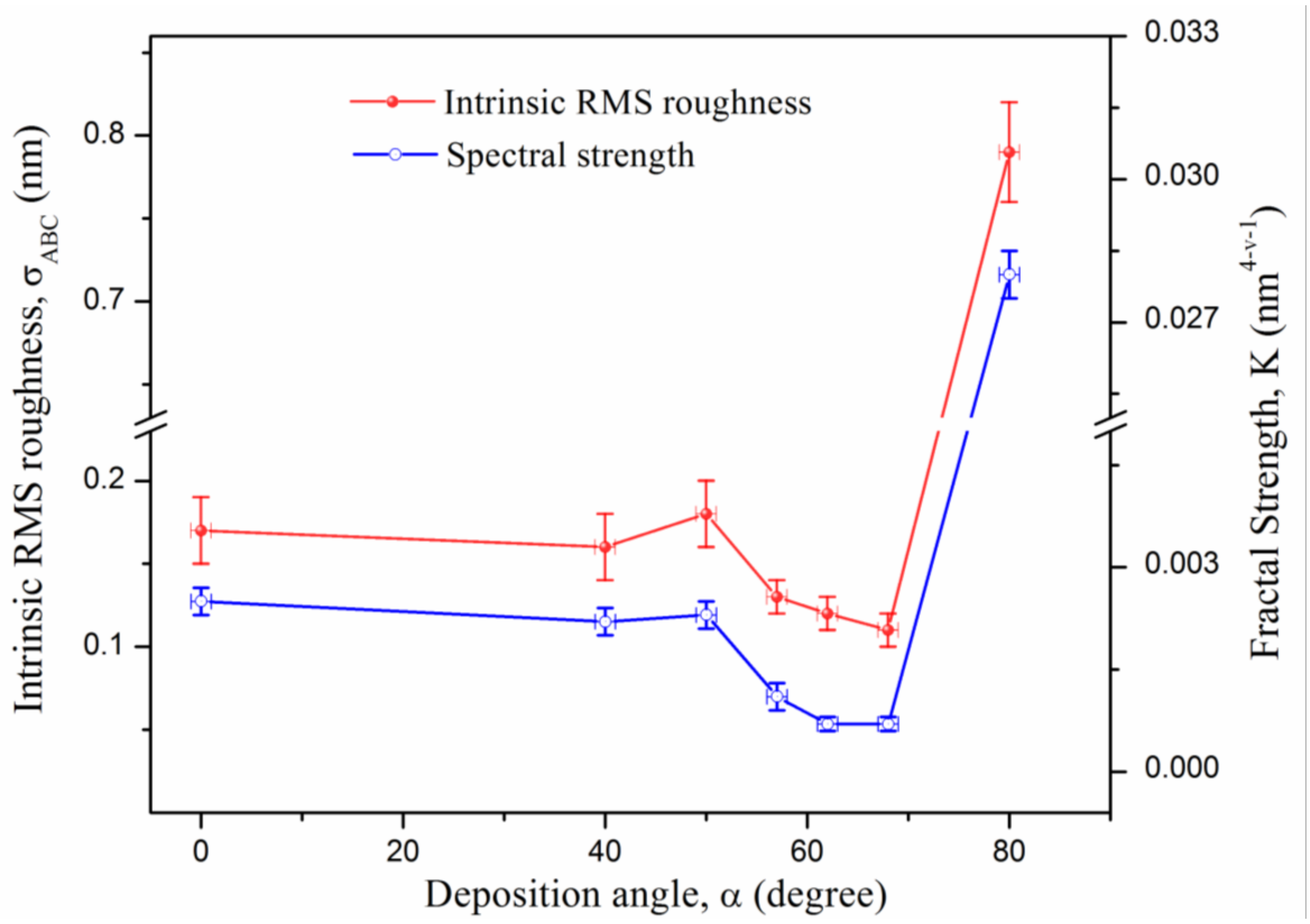

**Fig. 6:**

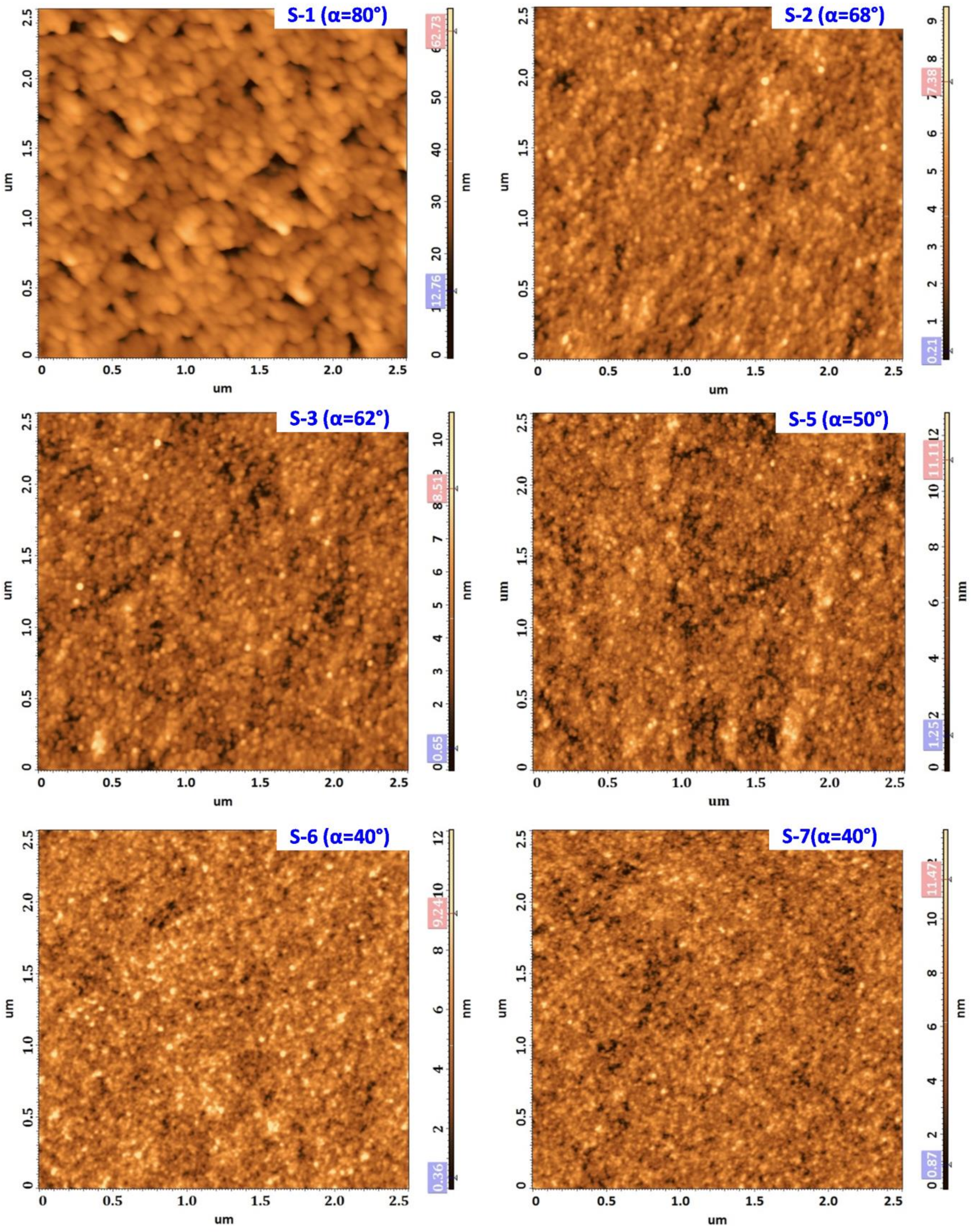

**Fig. 7(a):**

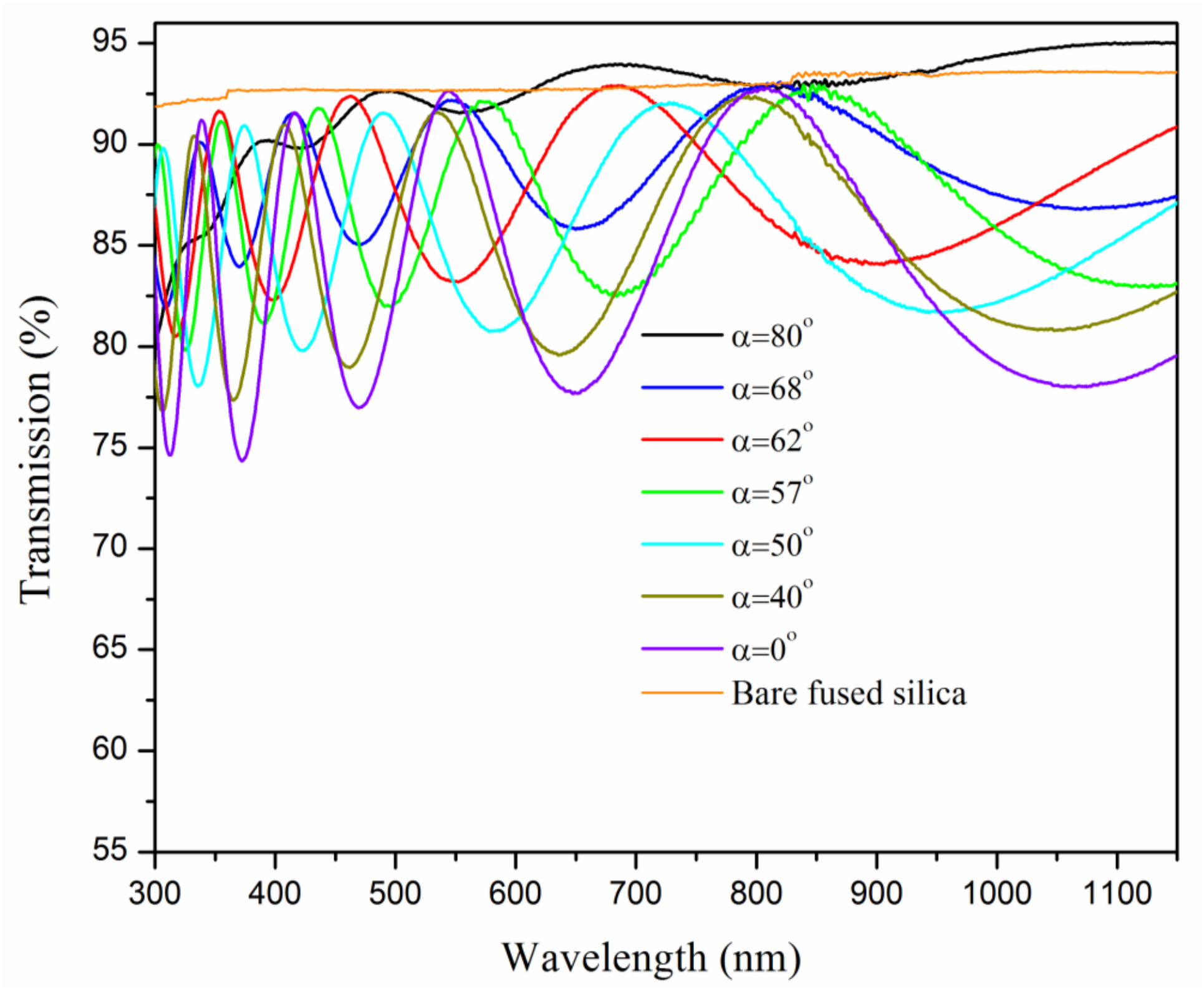

**Fig. 7(b):**

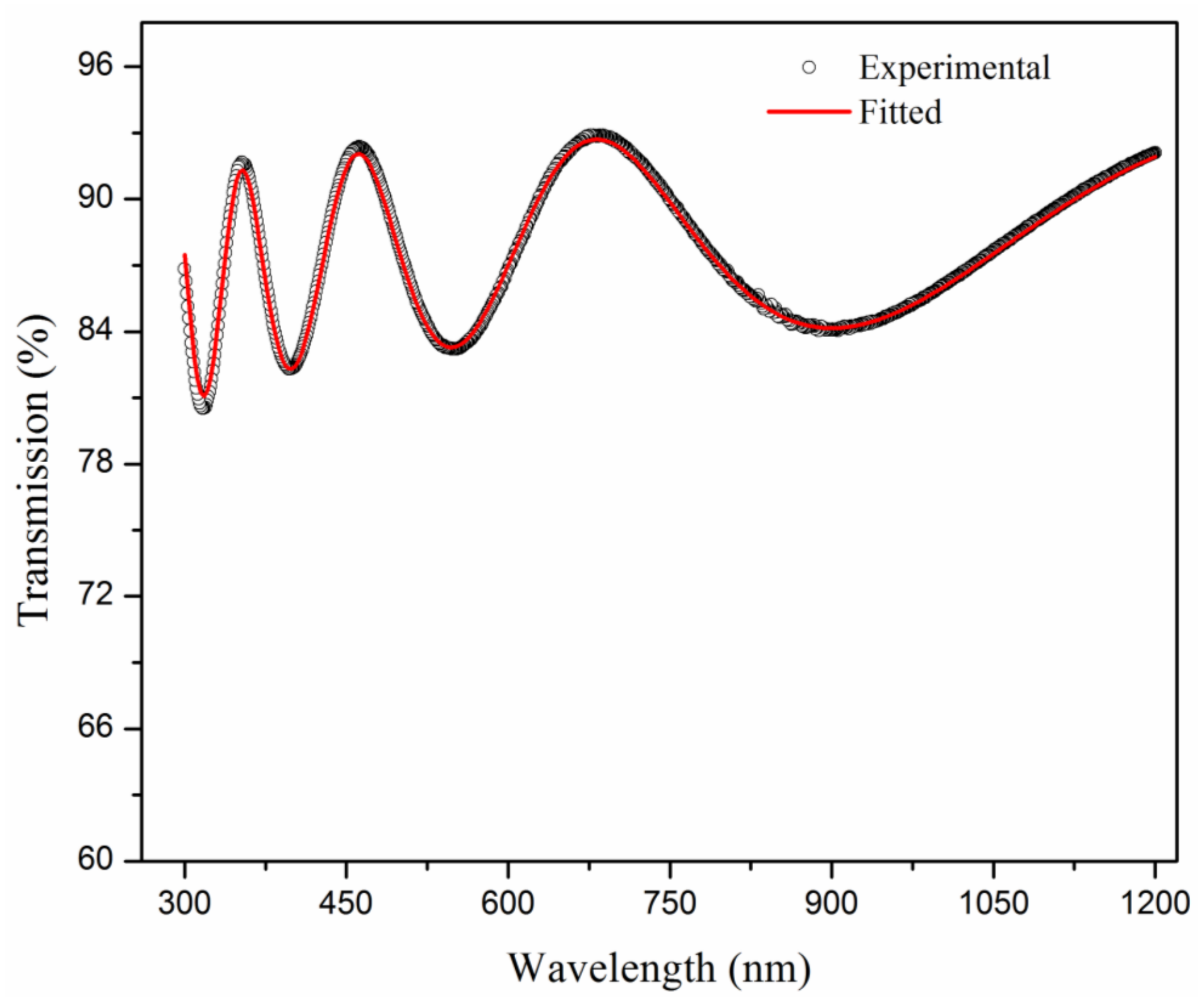

**Fig. 8:**

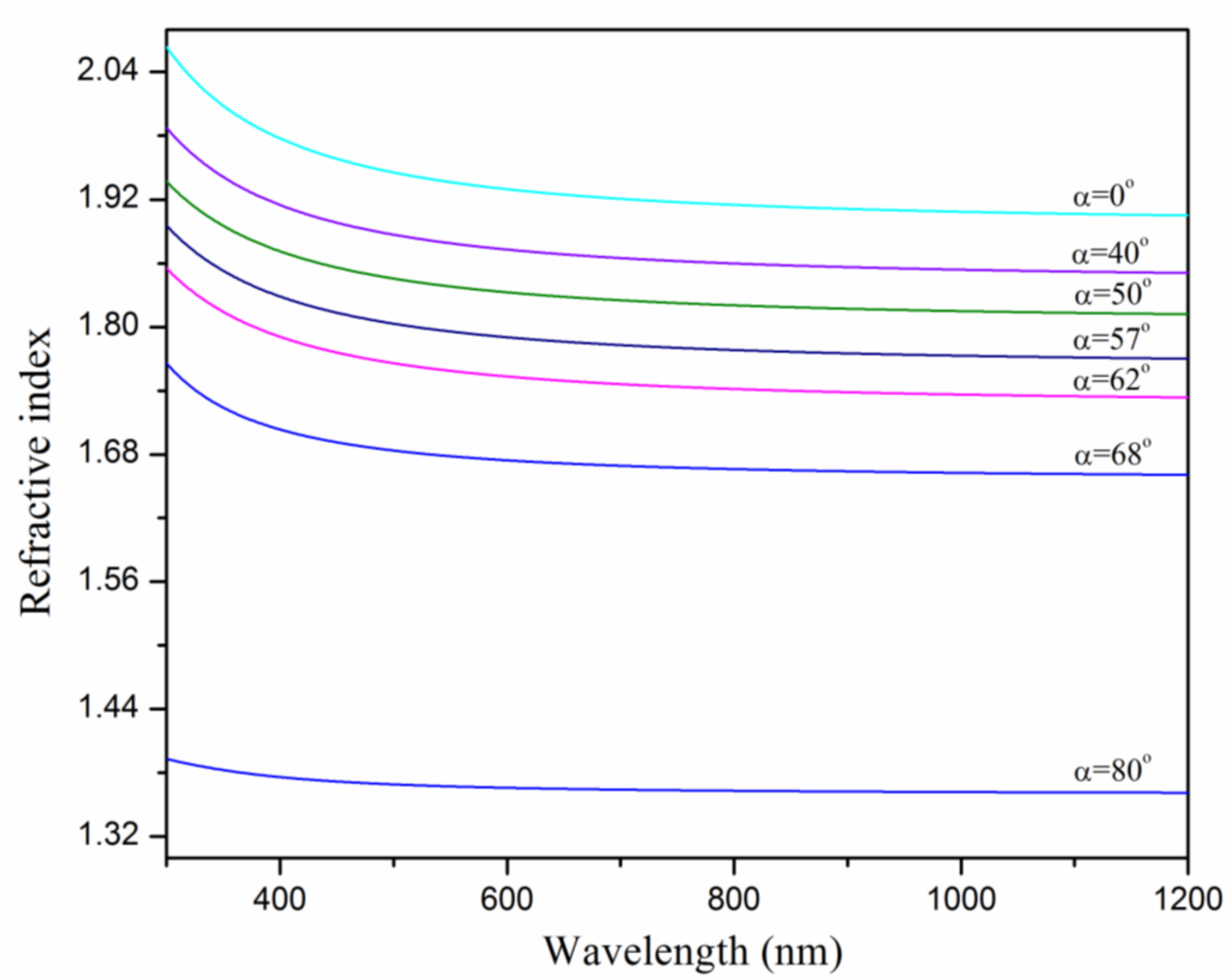

**Fig. 9:**

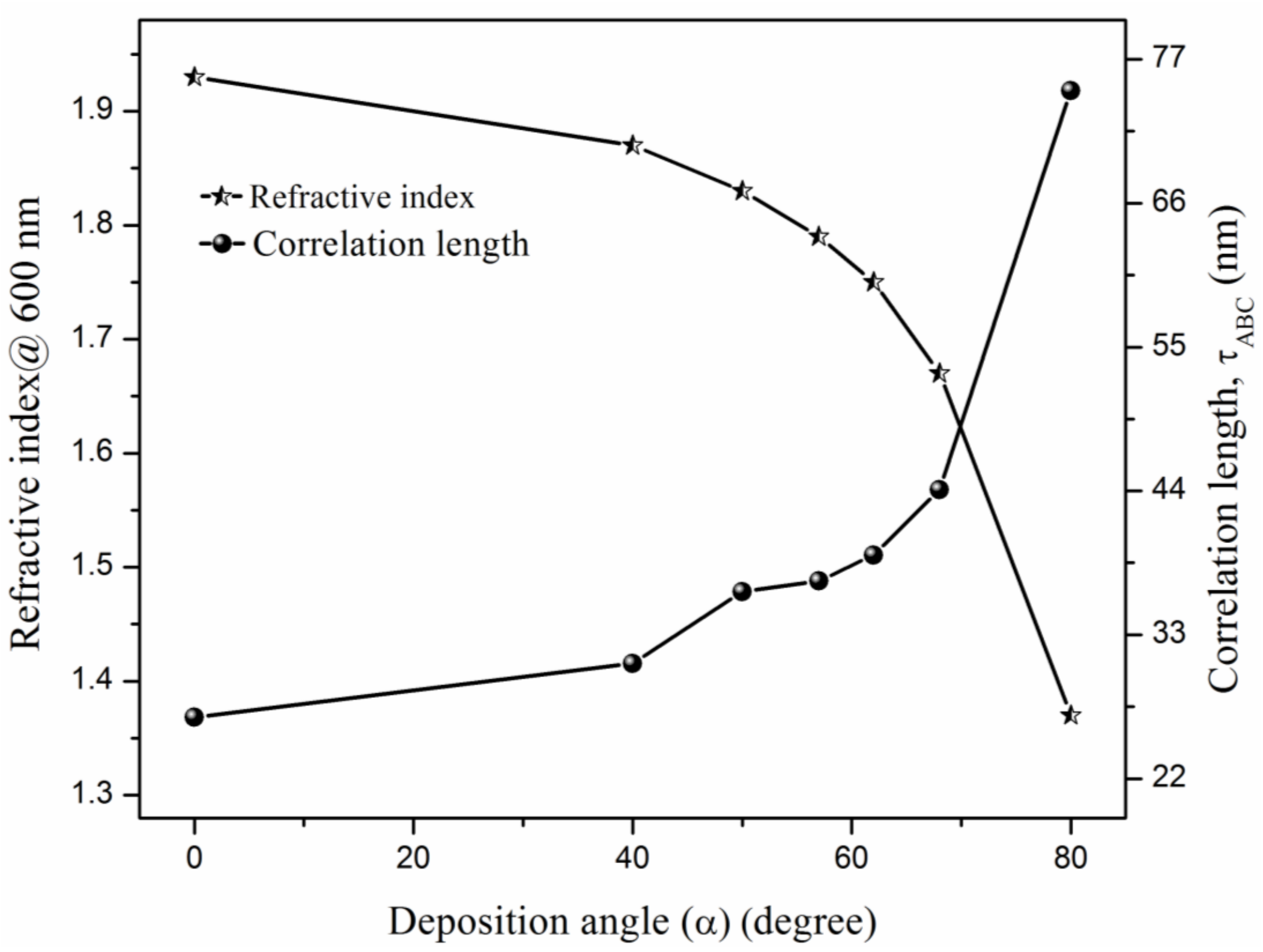

*Table-1: Fitting and derived morphological parameters from modelling of extended power spectral density function*

| *Sample name* | $\alpha$ *(degree)* | $\gamma$ | *K* *($nm^{4-\gamma-1}$) $x10^{-4}$* | *A* *($nm^4$)* | *B* *(nm)* | *C* | $\sigma_{ABC}$ *(nm)* | $\tau_{ABC}$ *(nm)* | $F_d$ | $\sigma_1$ *(nm) $x10^{-4}$* | $\tau_1$ *(nm)* | $f_{s1}$ *($nm^{-1}$) $x10^{-5}$* | $\sigma_2$ *(nm) $x10^{-4}$* | $\tau_2$ *(nm)* | $f_{s2}$ *($nm^{-1}$) $x10^{-4}$* |
|---|---|---|---|---|---|---|---|---|---|---|---|---|---|---|---|
| S-7 | 0 | 0.33 | 25 | 90 | 98 | 3.15 | 0.17+0.02 | 26.7+0.8 | 1.84 | 214 | 841 | 17 | 273 | 283 | 7 |
| S-6 | 40 | 0.36 | 22 | 110 | 105 | 3.35 | 0.16+0.02 | 30.8+0.8 | 1.82 | 170 | 870 | 22 | 210 | 250 | 6 |
| S-5 | 50 | 0.38 | 23 | 188 | 121 | 3.45 | 0.18+0.02 | 36.3+0.8 | 1.81 | 174 | 945 | 25 | 344 | 249 | 9 |
| S-4 | 57 | 0.40 | 11 | 102 | 125 | 3.45 | 0.13+0.01 | 37.1+.8 | 1.80 | 232 | 835 | 15 | 500 | 200 | 5 |
| S-3 | 62 | 0.48 | 7 | 95 | 130 | 3.50 | 0.12+0.01 | 39.1+0.8 | 1.76 | 167 | 948 | 21 | 176 | 380 | 9 |
| S-2 | 68 | 0.48 | 7 | 99 | 147 | 3.50 | 0.11+0.01 | 44.1+0.8 | 1.76 | 194 | 940 | 19 | 246 | 291 | 6 |
| S-1 | 80 | 0.48 | 280 | 15076 | 248 | 3.50 | 0.79+0.03 | 74.6+0.9 | 1.76 | 437 | 1510 | 24 | nil | nil | nil |